\documentclass[final,1p,times]{elsarticle} 
\usepackage{graphicx} 
\usepackage{amssymb} 
\usepackage{amsthm} 
\usepackage{lineno} 

\journal{Nuclear Physics A} 
\begin{document} 

\begin{frontmatter} 


\title{J/$\psi$ Production in $\it{p}$+$\it{p}$ and d+Au Collisions at $\sqrt{S_{NN}}$ = 200 GeV at STAR}

\author{Chris Perkins$^{a}$ for the STAR Collaboration}

\address[a]{UC Berkeley/Space Sciences Laboratory, 
Berkeley, CA 94720, USA}

\begin{abstract} 
We present analysis of J/$\psi$ production over the range $-1.0 < \eta < 4.2$ in $\it{p}$+$\it{p}$ and d+Au collisions using di-electron data taken during the 2008 run with the STAR experiment at Brookhaven National Laboratory.
STAR's unique forward capabilities, especially the Forward Meson Spectrometer electromagnetic calorimeter, allow us the possibility of investigating the intrinsic charm components of the proton wave function using high-$x_F$ forward particles produced in asymmetric partonic collisions.
Mid-rapidity measurements in d+Au collisions extend our understanding of the mechanisms underlying heavy quarkonium production and its transport through cold nuclear matter.

\end{abstract} 

\end{frontmatter} 


\section{Introduction and Motivation}

The J/$\psi$ production mechanism is expected to depend on rapidity at RHIC energies.
In the mid-rapidity region, it is expected to be dominated by gluon fusion but can also come from gluon or quark fragmentation or from feeddown from heavier mesons.
At forward rapidity, little is known about J/$\psi$ production.
It has been hypothesized that in this region the dominant production mechanism could come from intrinsic heavy flavor components in the proton wavefunction~\cite{brodsky-2009-807}.

While the light sea-quark components of the proton wavefunction have been measured over a wide range of proton momentum fraction $\it{x}$, little is known about the intrinsic components of heavier quarks, especially at large $\it{x}$.
The QCD heavy flavor fluctuations can manifest themselves as high Feynman $\it{x}$ ($x_F$ = 2$p_{L}$/$\sqrt{s}$) quarkonium states, requiring their observation at forward rapdities at RHIC energies.
A measurement of high-$x_{F}$ quarkonium could provide important clues about intrinsic heavy flavor in the proton with consequences on a possible diffractive Higgs production measurement at the LHC~\cite{brodsky-2009-807}.

Measurements at mid-rapidity in d+Au collisions may shed light on whether the J/$\psi$ is produced in a color singlet or color octet state and the relative production contributions from gluon fusion, quark and gluon fragmentation, and decay feeddown.
Mid-rapidity measurements also explore the heavy flavor transport mechanisms through the Quark Gluon Plasma providing a better understanding of color screening and suppression in the medium.
Both low-$p_T$ and high-$p_T$ measurements at mid-rapidity are critical in d+Au collisions in order to disentangle cold nuclear effects from hot nuclear effects to better understand heavier systems such as Au+Au or Cu+Cu.

We report observations of J/$\psi$ at STAR at both forward and mid-rapidity using data collected during RHIC Run 8 at $\sqrt{s_{NN}}$ = 200 GeV.
The mid-rapidity measurements were made using d+Au collisions and include measurements of $R_{dAu}$ and J/$\psi$-hadron azimuthal correlations.
The forward rapidity, high-$x_F$ J/$\psi$ observations were made in $\it{p}$+$\it{p}$ collisions and are the first made at collider energies $>$ 62 GeV.
The reported measurements observe the J/$\psi$ through its decay channel J/$\psi$ $\rightarrow$ $e^+$ + $e^{-}$ by reconstructing the invariant mass of the $e^{+}$ $e^{-}$ pair.
An integrated sampled luminosity of 50 nb$^{-1}$ from d+Au and 7.8 pb$^{-1}$ from $\it{p}$+$\it{p}$ were collected for this analysis.

\section{Experimental Setup}
The reported measurements were made using the Solenoidal Tracker at RHIC (STAR).

The forward rapidity measurement was made using the Forward Meson Spectrometer (FMS), an electromagnetic calorimeter located approximately 7.5 meters to the west of the STAR interaction point and the Beam-Beam Counters (BBC) for additional offline event selection.
The FMS provides full azimuthal coverage over a pseudorapidity range of approximately $2.5 < \eta < 4$.
The forward rapidity data sample was collected in $\it{p}$+$\it{p}$ collisions using a high-tower trigger on the FMS cells with a BBC minimum bias condition imposed offline.

The mid-rapidity measurements were made using the Time Projection Chamber (TPC), the Barrel Electromagnetic Calorimeter (BEMC), and the Barrel Shower Max Detector (BSMD).
The high-$p_{T}$ measurement was based on the data sample collected by triggering on high towers in the BEMC, providing a fast trigger while still yielding a rich electron sample.
The electrons in this sample were identified using dE/dx cuts on TPC data and the BEMC and BSMD to provide additional electron/hadron separation.
The low-$p_{T}$ measurement was based on the data sample collected with a minimum bias trigger.
Electrons were identified in this sample using a dE/dx cut on TPC data and a p/E cut on BEMC data.

\section{Mid-Rapidity J/$\psi$ Measurements (d+Au)}

\begin{figure}[ht]
\centering
\includegraphics[width=0.45\textwidth]{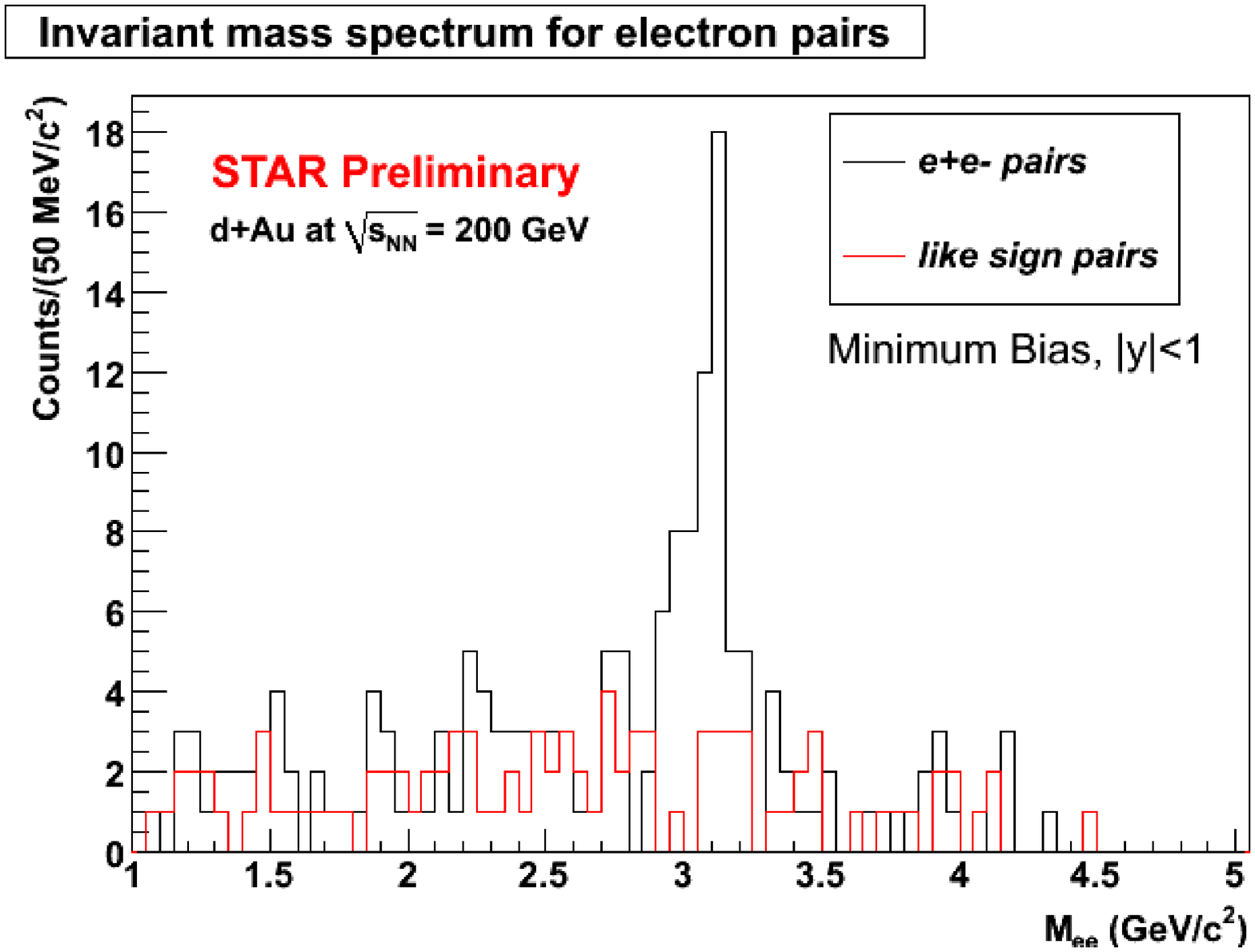}
\includegraphics[width=0.45\textwidth]{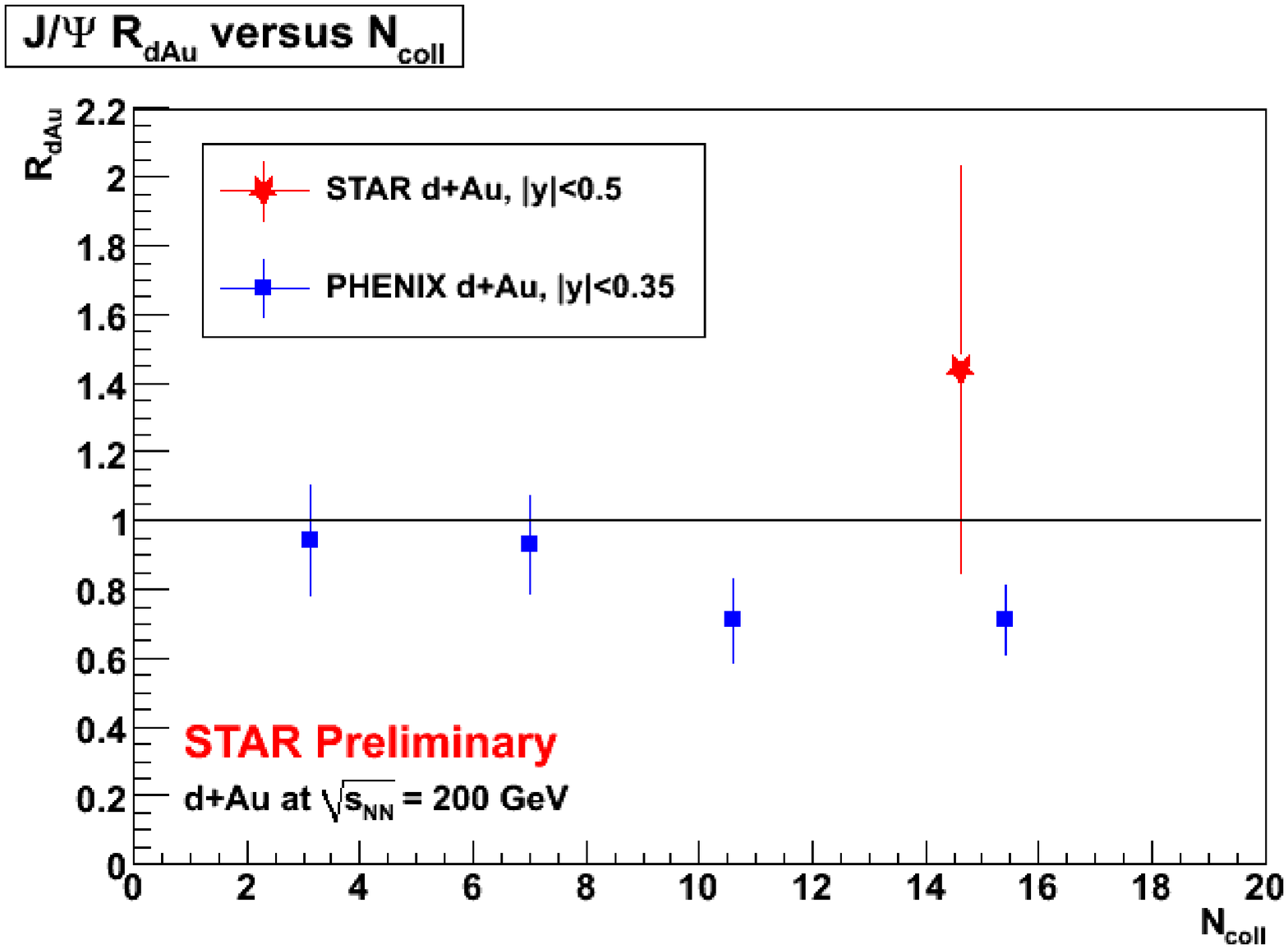}
\caption{(a) Low-$p_T$ invariant dielectron mass distribution in d+Au collisions, for opposite-sign (black curve) and like-sign (red curve) pairs from data. Sample includes all minbias triggered events over a rapidity range $-1 < y < 1$. (b) Low-$p_T$ J/$\psi$ $R_{dAu}$ using only the most central events (0-20\%) over a rapidity range $-0.5 < y < 0.5$. }
\label{low_pt}
\end{figure}

The reconstructed invariant mass of low-$p_{T}$ J/$\psi$ is shown in figure~\ref{low_pt}(a).
This measurement used the full minimum bias data sample over a rapidity range $|y| < 1$.
The removal of material near the interaction region prior to the 2008 run has resulted in a much better signal to background ratio than previous low-$p_{T}$ J/$\psi$ measurements at STAR.
An efficiency corrected d+Au yield scaled by the number of binary collisions, $R_{dAu}$ = $\frac{dN^{dAu}/dy}{N_{coll}dN^{pp}/dy}$, is shown in figure~\ref{low_pt}(b).
This measurement used the most central events (0-20\%) over a rapidity range $|y| < 0.5$.
We report a measurement of $R_{dAu}$ = 1.4 $\pm$ 0.6 (stat. errors only), consistent with unity, as previously reported.

\begin{figure}
\centering
\includegraphics[width=0.32\textwidth]{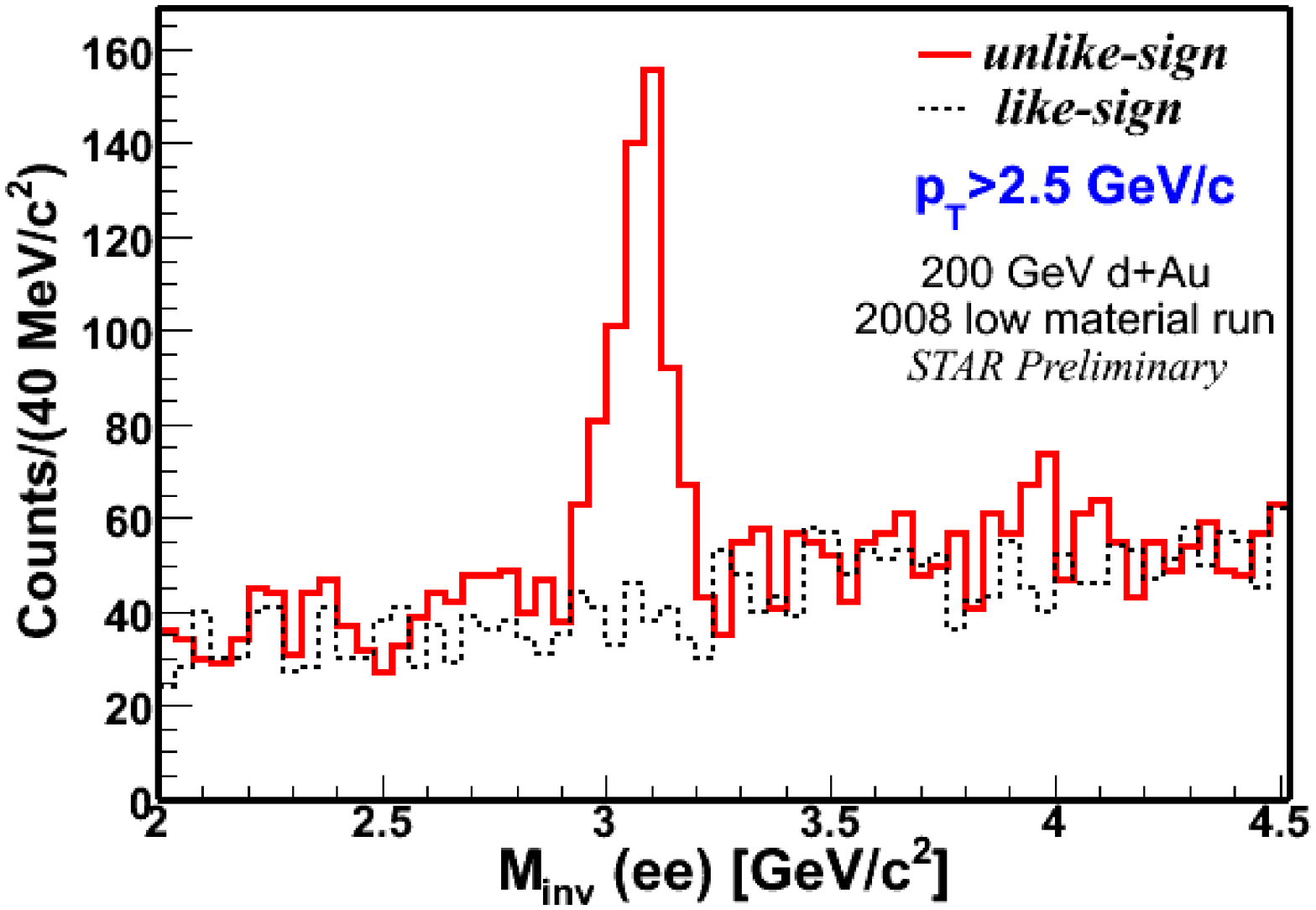}
\includegraphics[width=0.32\textwidth]{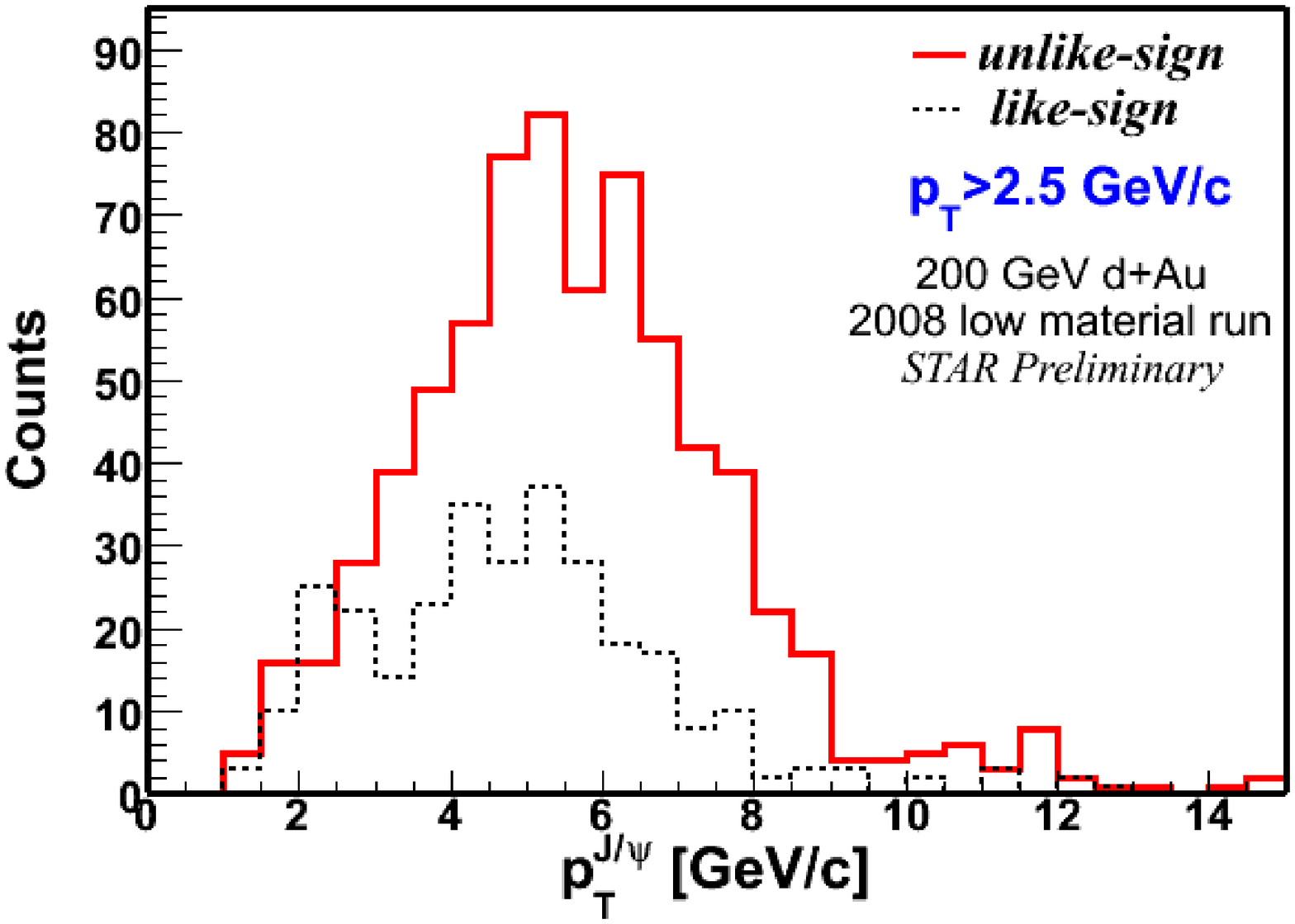}
\includegraphics[width=0.32\textwidth]{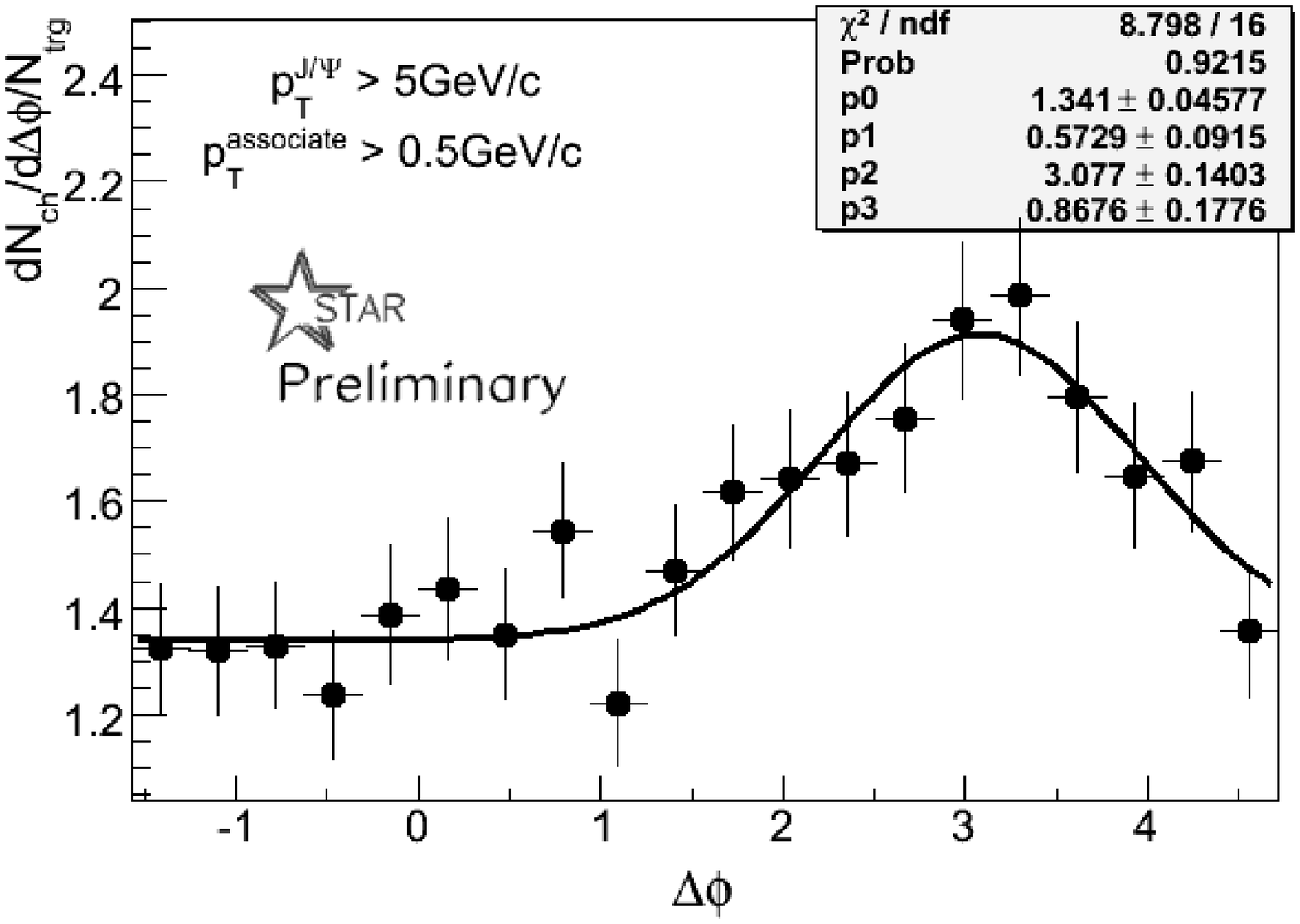}
\caption{(a) High-$p_T$ invariant dielectron mass distribution in d+Au collisions, for opposite-sign (black curve) and like-sign (red curve) pairs. (b) Triggered $p_T$ distribution of high-$p_T$ J/$\psi$. (c) High-$p_T$ J/$\psi$-hadron azimuthal correlations.}
\label{high_pt}
\end{figure}

The reconstructed invariant mass of high-$p_{T}$ J/$\psi$ is shown in figure~\ref{high_pt}(a).
We report an observation of J/$\psi$ with a 10$\sigma$ signficance.
A triggered $p_{T}$ spectrum (not yet efficiency corrected) for these reconstructed J/$\psi$s is shown in figure~\ref{high_pt}(b).
Monte Carlo studies have shown~\cite{jpsi_feeddown_1} ~\cite{jpsi_feeddown_2} that J/$\psi$ production coming from the feeddown of $\chi_c$ give no near-side azimuthal correlations between J/$\psi$ and associated hadrons while feeddown coming from B meson decays leads to a strong near-side azimuthal correlation.
We see no significant near-side, high-$p_T$ J/$\psi$-hadron azimuthal correlations in d+Au collisions (See Figure~\ref{high_pt}(c)) which may help us constrain the feeddown contribution to J/$\psi$ production from B mesons.

\section{Forward Rapidity J/$\psi$ Measurements ($\it{p}$+$\it{p}$)}
Because there is no tracking detector that overlaps with the FMS, we cannot distinguish the charge sign of the J/$\psi$ decay electrons and must therefore rely on simulations for the background shape in the region of the J/$\psi$ mass.
In addition to performing a minimum bias simulation to verify our reconstruction procedure and resolutions, we performed a high-mass filtered PYTHIA + GEANT simulation to generate the background shape.

\begin{figure}
\centering
\includegraphics[width=0.32\textwidth]{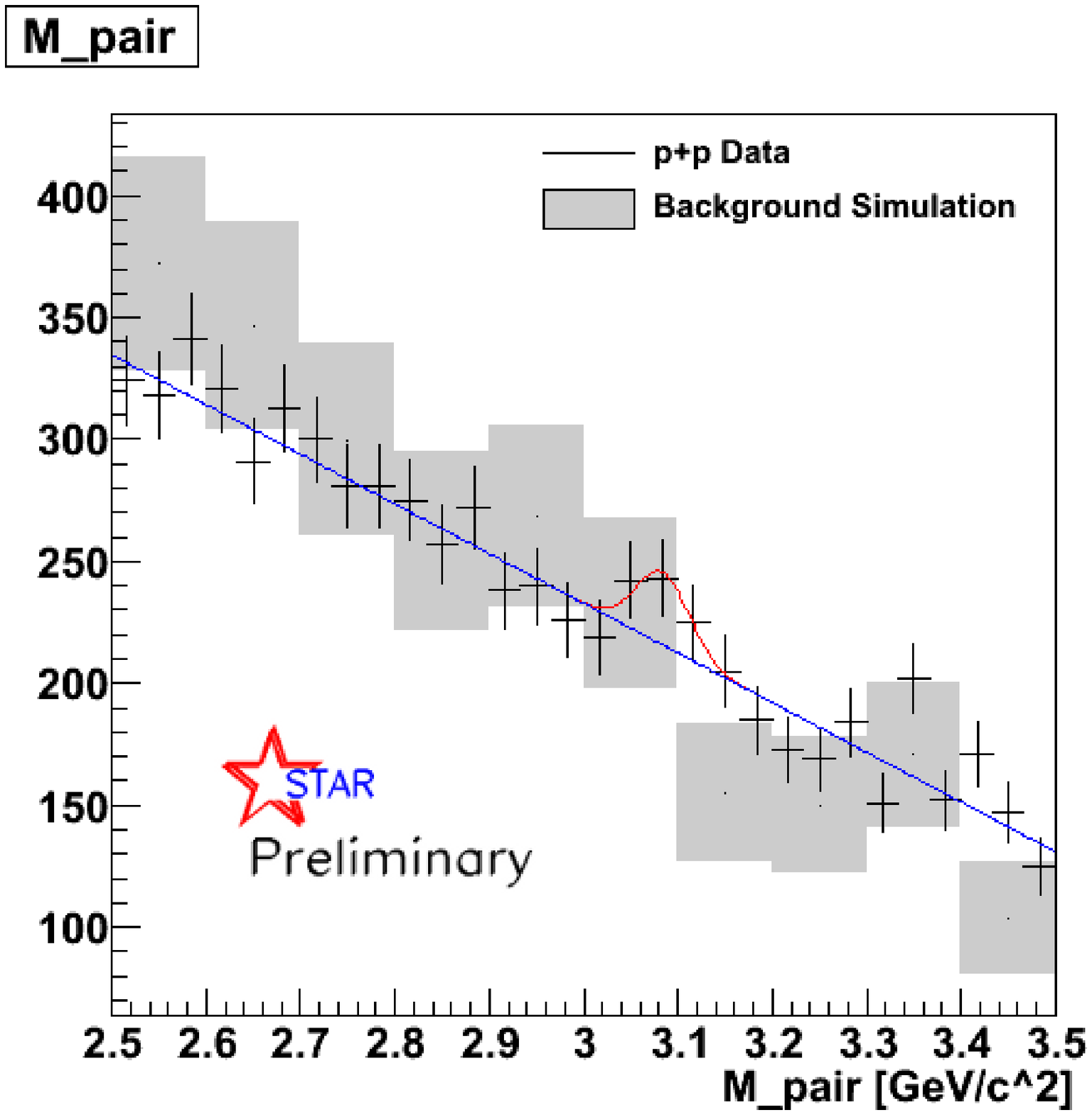}
\includegraphics[width=0.32\textwidth]{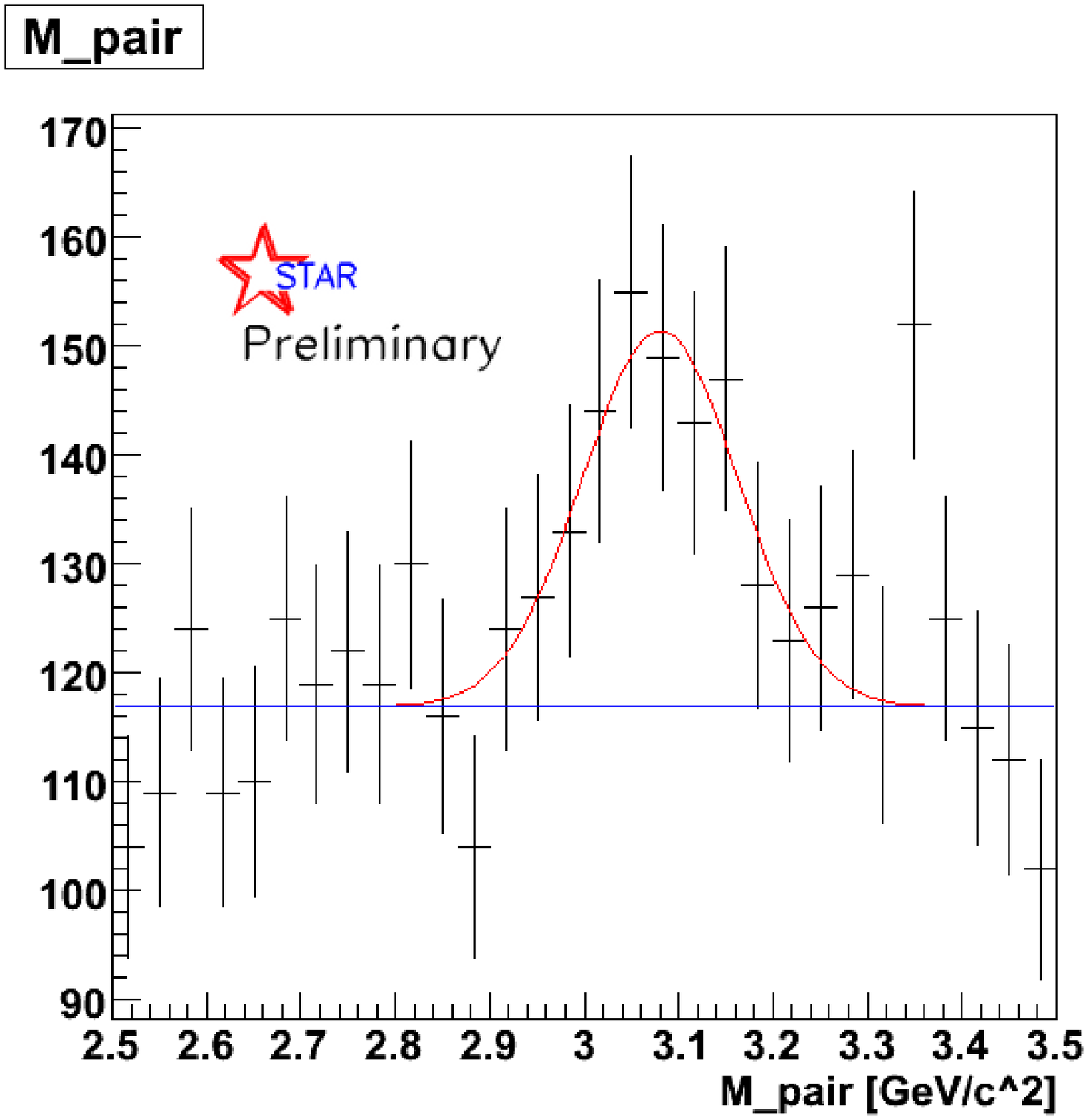}
\includegraphics[width=0.32\textwidth]{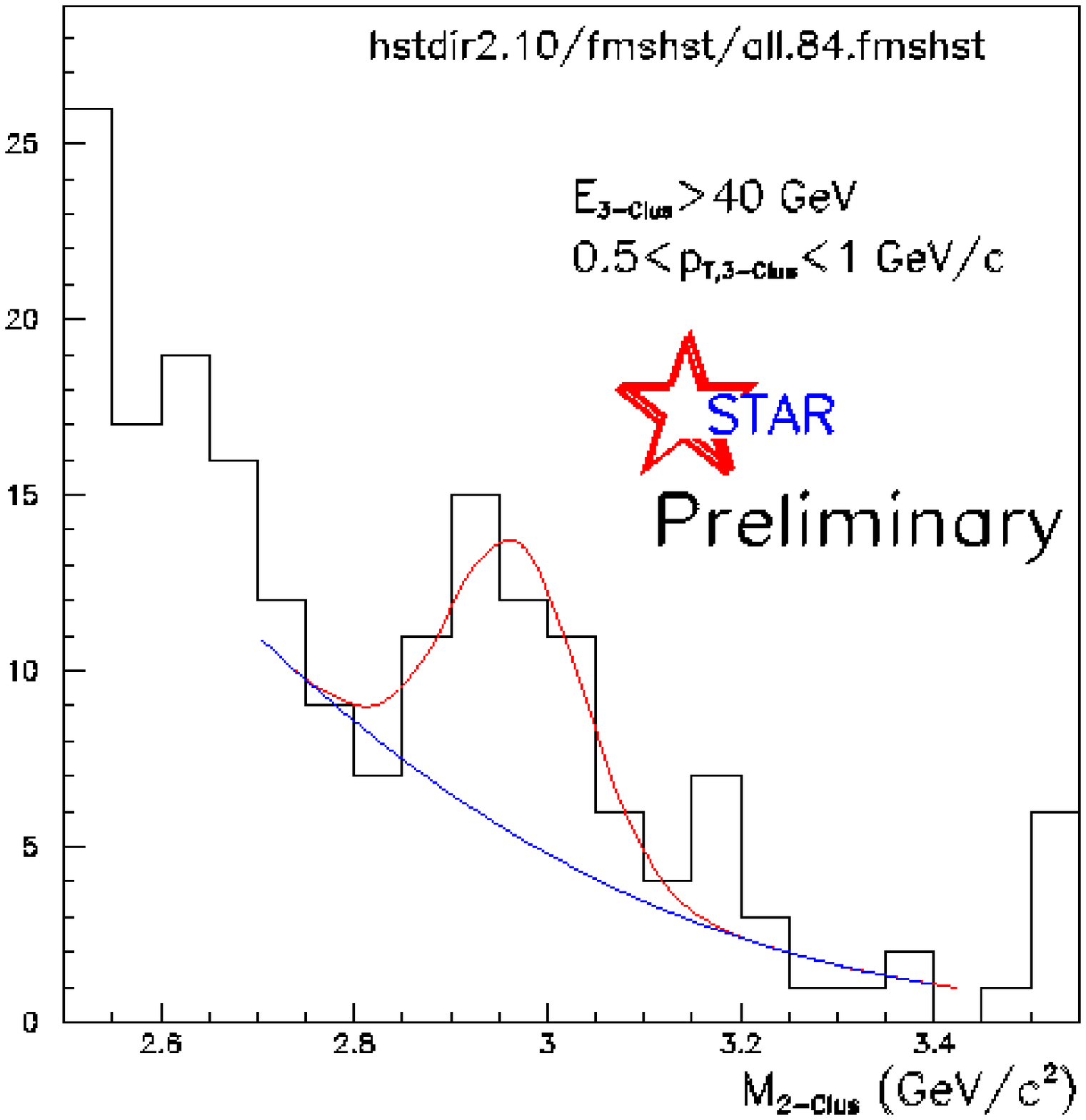}
\caption{(a) Reconstructed invariant mass at forward rapidity in $\it{p}$+$\it{p}$ collisions with original cuts for 2-cluster analysis. Errors bars on data points are statistical only. Blue curve shows fitted background.  Red curve shows fitted J/$\psi$ signal. Gray bands show simulated background. (b) Reconstructed invariant mass at forward rapidity for 2-cluster analysis with additional cuts on cluster $p_T$. (c) Reconstructed invariant mass of pair associated with J/$\psi$ in 3-cluster analysis. }
\label{m_inv}
\end{figure}

The reconstructed 2-cluster mass is shown in figure~\ref{m_inv}(a).
This plot requires that the pair energy is greater than 60.0 GeV, the energy sharing, Z = $\frac{E_{e^+} - E_{e^-}}{E_{e^+} + E_{e^-}}$, to be less than 0.7, and that clusters are isolated by R=$\sqrt{(\Delta\eta)^2 + (\Delta\phi)^2} > 0.5$.
The isolation cut suppresses combinatorial background from lower mass particles.
The energy sharing cut focuses the analysis on relatively symmetric decays to ensure that most of the J/$\psi$ decay products will both fall within the acceptance of the FMS.
The signal was fitted with a gaussian plus a linear polynomial, resulting in a fit with a significance of 2.1$\sigma$.
The background simulation was normalized to the integral of the data in this mass region.
Even with the PYTHIA filter applied, background simulation statistics are still fairly low but show that the shape of the background is generally correct.
An improved background shape with more statistics is in progress.

A further cut was applied to the $p_T$ of individual clusters to filter those clusters coming from the large background near the beampipe.
The reconstructed 2-cluster invariant mass with this new cut and a slightly lower isolation radius cut is shown in figure~\ref{m_inv}(b).
This data was fitted with a gaussian plus an offset, resulting in a fit with a significance of 4.5$\sigma$.
The mean of the gaussian fit is very close to the expected J/$\psi$ mass.
Incomplete field effect simulations and energy dependent gain calibrations can account for a fit width that is lower than expectations from simulations.

We also performed a 3-cluster analysis that observes J/$\psi$ through its feeddown from $\chi_c$ ($\chi_c \rightarrow J/\psi + \gamma \rightarrow e^+ + e^- + \gamma$).
This technique was motivated by an analysis of $\omega \rightarrow \pi^0 + \gamma$ as shown in~\cite{agordon} and was also used by the COMPASS experiment in looking at $D \rightarrow K + \pi$ in D* resonance decays~\cite{compass}.
For each group of 3 clusters within an event, we associate the pair with reconstructed mass closest to the accepted J/$\psi$ mass (3.097 GeV/$c^2$) with the J/$\psi$ and the remaining cluster with the $\gamma$.
Mass plots of the pair chosen to be the J/$\psi$ compared with mass plots of the other cluster combinations indicate that we are correctly identifying the J/$\psi$.

The reconstructed invariant mass of the pair associated with the J/$\psi$ from the 3-cluster analysis is shown in figure~\ref{m_inv}(c).
The background and signal were fitted with separate gaussians resulting in a fit with a signficance of 2.9$\sigma$.
We found that the significance of the fit depends highly on the background model used.
A realistic simulation of this background is still in progress but the reported significance of 2.9$\sigma$ is a conservative estimate.
The mean and sigma of the gaussian fit are also lower than expected, as was found in the 2-cluster analysis.
This observation could suffer from the same field effects and energy calibration issues described previously.

\section{Conclusions and Outlook}
At mid-rapidity, we report a significant signal for both high-$p_T$ and low-$p_T$ J/$\psi$ in d+Au collisions.
At low-$p_T$ we also report a measurement of $R_{dAu}$ = 1.4 $\pm$ 0.6 (stat.), consistent with scaling by the number of binary collisions.
At high-$p_T$ we see no significant near-side J/$\psi$-hadron correlations, which can constrain feeddown from B mesons.

At forward rapidities, we report the observation of a J/$\psi$ signal for $\it{p}$+$\it{p}$ collisions using both our 2-cluster and 3-cluster techniques.
In using the 3-cluster technique, we not only report a significant J/$\psi$ signal but also provide a first look at high-$x_F$ $\chi_c \rightarrow J/\psi + \gamma$ feeddown.
These observations are the first observations of high-$x_F$ J/$\psi$ at $\sqrt{s} > 62$ GeV.
In future RHIC runs at $\sqrt{s}$ = 500 GeV we also plan to look for high-$x_{F}$ Upsilons which, with the high-$x_{F}$ J/$\psi$, could provide further insight in the understanding of the virtual heavy quark content within the proton.



\end{document}